\documentclass{PoS}

\usepackage{amssymb}
\usepackage{amsthm}
\usepackage{mathrsfs}
\usepackage{amsmath}

\usepackage{tabularx}
    \newcolumntype{L}{>{\raggedright\arraybackslash}X}
    
\title{Multi-messenger Gravitational-Wave + High-Energy Neutrino Searches with LIGO, Virgo, and IceCube}

\ShortTitle{Gravitational-Wave + High-Energy Neutrino Searches}

\author{The IceCube Collaboration\footnote{For collaboration list, see PoS(ICRC2019) 1177.}\\
{\itshape \href{http://icecube.wisc.edu/collaboration/authors/icrc19_icecube}{http://icecube.wisc.edu/collaboration/authors/icrc19\_icecube}}\\
E-mail: \email{azadeh.keivani@icecube.wisc.edu}
}

\abstract{

Multi-messenger searches for gravitational waves and high-energy neutrinos provide important insights into the dynamics of and particle acceleration by black holes and neutron stars. With LIGO's third observing period (O3), the number of gravitational wave detections has been substantially increased. The rapid identification of joint signals is crucial for electromagnetic follow-up observations of transient emission that is only detectable for short periods of time. High-energy neutrino direction can be reconstructed to sub-degree precision, making a joint detection far better localized than a standalone gravitational-wave signal. We present the latest sensitivity of joint searches
and discuss the Low-Latency Algorithm for Multi-messenger Astrophysics (LLAMA) that combines LIGO/Virgo gravitational-wave candidates and searches in low-latency for coincident high-energy neutrinos from the IceCube Neutrino Observatory. We will further discuss future prospects of joint searches from the perspective of better understanding the interaction of relativistic and sub-relativistic outflows from binary neutron star mergers.\\

\vspace{4mm}
{\bfseries Corresponding authors: }\speaker{Azadeh Keivani}$^{1}$,  {Do\u{g}a Veske}$^{1}$, {Stefan Countryman}$^{1}$, {Imre Bartos}$^{2}$, {K.~Rainer Corley}$^{1}$, {Zsuzsa Marka}$^{1}$, {Szabolcs Marka}$^{1}$\\
{$^{1}$ \itshape Columbia University}\\
{$^{2}$ \itshape University of Florida}\\

}

\FullConference{36th International Cosmic Ray Conference -ICRC2019-\\
		July 24th - August 1st, 2019\\
		Madison, WI, U.S.A.}

\begin{document}

\section{Introduction}\label{sec:intro}

The IceCube Neutrino Observatory has identified and confirmed a quasi-diffuse flux of high-energy neutrinos (HENs) over years \cite{Aartsen:2013jdh,PRL2014,chaack2017}.
In addition, IceCube detected a HEN (IceCube-170922A) in 2017 through its realtime program~\cite{IC_realtime}, which for the first time found to be correlated (at a 3$\sigma$ significance level) with a flaring blazar in an extensive multi-messenger observational campaign~\cite{ic170922a}.
An archival search further showed an additional neutrino flare candidate from the direction of the blazar in 2014-2015~\cite{ic170922a-2}.
On the other hand, LIGO and Virgo gravitational wave (GW) observatories have so far detected several GW events~\cite{2018arXiv181112907T}. 
One of these events, GW170817, was observed in company of broad-band electromagnetic (EM) emission and resulted in comprehensive study of a binary neutron star (BNS) system~\cite{gw170817}. 
Although GWs and HENs have been observed in coincidence with EM emissions, there has been no astrophysical source yet observed emitting both GWs and HENs~\cite{2019ApJ...870..134A}. 
However, we expect to find such sources in near future considering the increasing rate of GW detection due to improving and upgrading the detectors~\cite{shigeo2018}. 

The most promising jointly emitting GW+HEN source candidate is short gamma-ray burst (sGRB) produced by its progenitor, BNS merger. 
Any gravitational wave source with relativistic outflows capable of producing HENs through hadronic interactions could in principle be a GW+HEN source candidate. 
This includes neutron star$-$black hole mergers, core collapse supernovae, gamma-ray bursts, and soft gamma repeaters (see references within~\cite{gwhenpipeline}).

We have developed an online data analysis pipeline to search for multi-messenger sources of GWs and HENs which was mainly adapted from an original work of~\cite{2012PhRvD..85j3004B}. 
The Low-Latency Algorithm for Multi-messenger Astrophysics (LLAMA;~\cite{gwhenpipeline,gwhenmethod}) searches for significant coincident GW+HEN events and broadcasts the search results through gamma-ray coordinates network (GCN;~\cite{gcn}) circulars to the astrophysical community. 
Such alerts would usually have far better localization compared to the GW events alone due to smaller angular errors of neutrinos, making them more suitable for EM follow-up observations. 
Finding even one such alert can revolutionize our understanding of their astrophysical source. 

There are currently two different analyses in IceCube searching for GW+HEN events. 
Here we focus on describing the LLAMA search which relies on material discussed in depth in~\cite{gwhenpipeline,gwhenmethod}.
The other analysis is discussed in \cite{Hussain:2019icrc_gw}.
The IceCube GCN circulars on GW+HEN events report the results of both of these analyses for any GW public alert. 

\section{Pipeline}\label{sec:pipeline}

\begin{figure}
\begin{center}
\begin{tabular}{cc}
    \includegraphics[width=0.47\linewidth, trim=0cm 0cm 0cm 0.5cm]{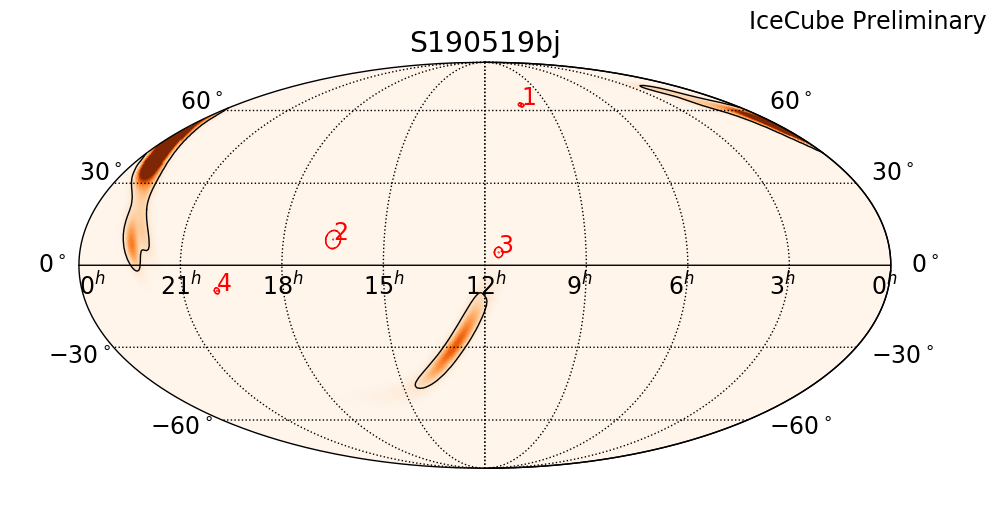} &
    \includegraphics[width=0.47\linewidth, trim=0cm 0cm 0cm 0.5cm]{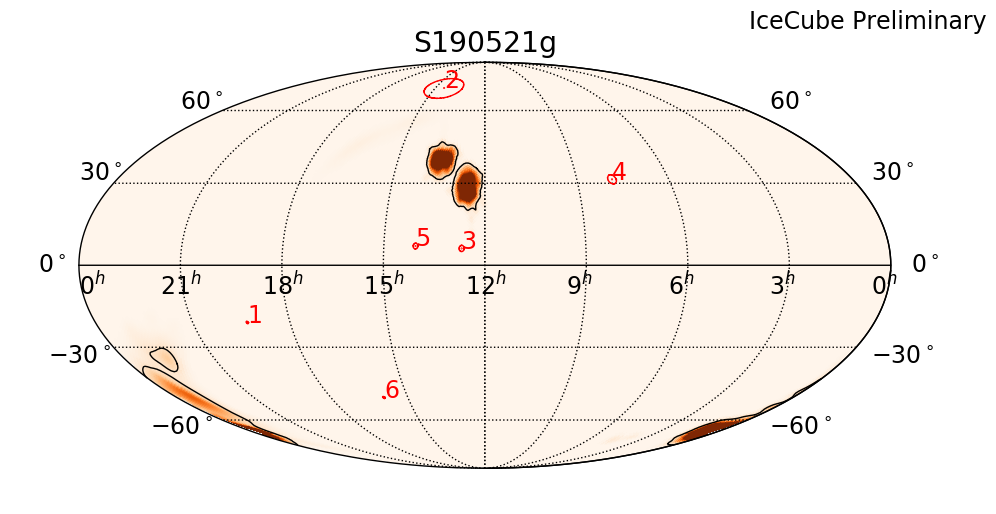} \\
    \includegraphics[width=0.47\linewidth, trim=0cm 0cm 0cm 0.5cm]{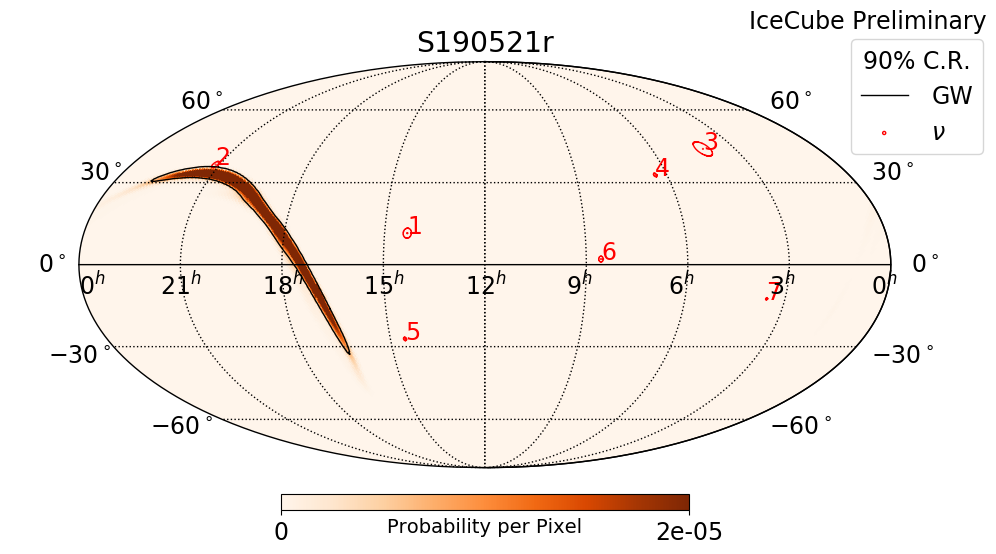} &
    \includegraphics[width=0.47\linewidth, trim=0cm 0cm 0cm 0.5cm]{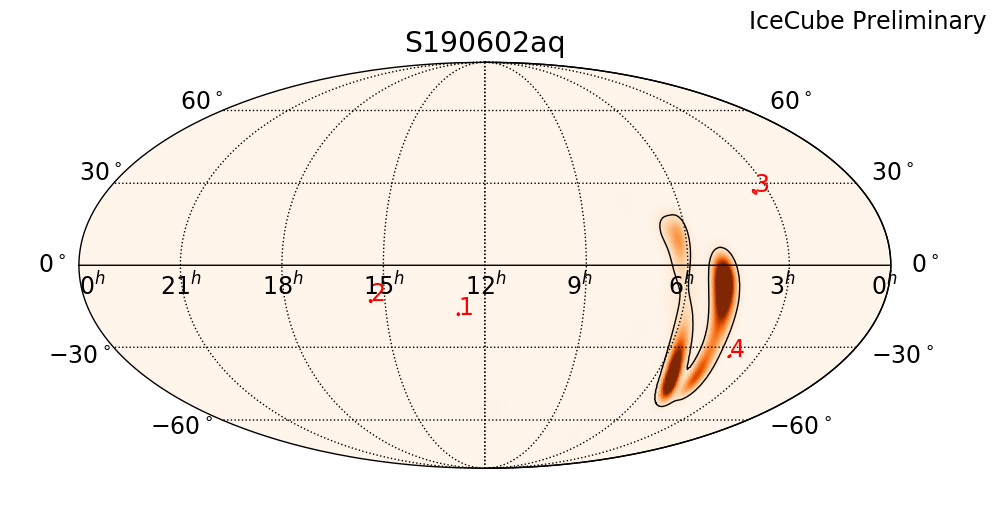}
\end{tabular}
\end{center}
\caption{Joint GW+HEN skymaps for the 14 detected GW events since the beginning of O3. Neutrinos temporally correlated with any of the GW events ($\pm$ 500 seconds of the GW trigger time) are overlaid on each skymap. None of the neutrinos made a significant coincidence in any of the searches. The 90\% confidence regions of GWs and HENs are plotted in black and red contours, respectively.  The maps are preliminary and plotted in celestial (equatorial, J2000) coordinates.}
\label{fig:skymap}
\end{figure}

LLAMA triggers on LIGO/Virgo public alerts and pulls data from GraceDB\footnote{https://gracedb.ligo.org}. 
For any new GW public alert, an event directory on LLAMA server's filesystem is generated. 
It then pulls data from IceCube gamma-ray follow-up (GFU)~\cite{icgfu2016} stream as data become available and run the analysis described in Sec.~\ref{sec:method} to provide a joint skymap and a significance value for the GW+HEN alert and send out the information to IceCube realtime oversight committee (ROC) for distribution through a GCN circular. 
This information is also uploaded to GraceDB and the IceCube Slack. 
The pipeline's analysis steps are discussed in detail in~\cite{gwhenpipeline}.
The pipeline repeatedly tries to download data and for any GW trigger, searches for HENs in a time window of 1000~s of the GW detected time~\cite{2011APh....35....1B}. 
The analysis is complete in a few seconds and skymaps such as the ones in Figure~\ref{fig:skymap} are immediately produced.
The skymaps are for internal use in IceCube and the information about individual neutrinos will be provided in the GCN circular only in case of a significant coincident GW+HEN event (see Sec.~\ref{sec:method}).


\begin{center}
\begin{tabular}{cc}
    \includegraphics[width=0.47\textwidth, trim=0cm 0.5cm 0cm 0cm]{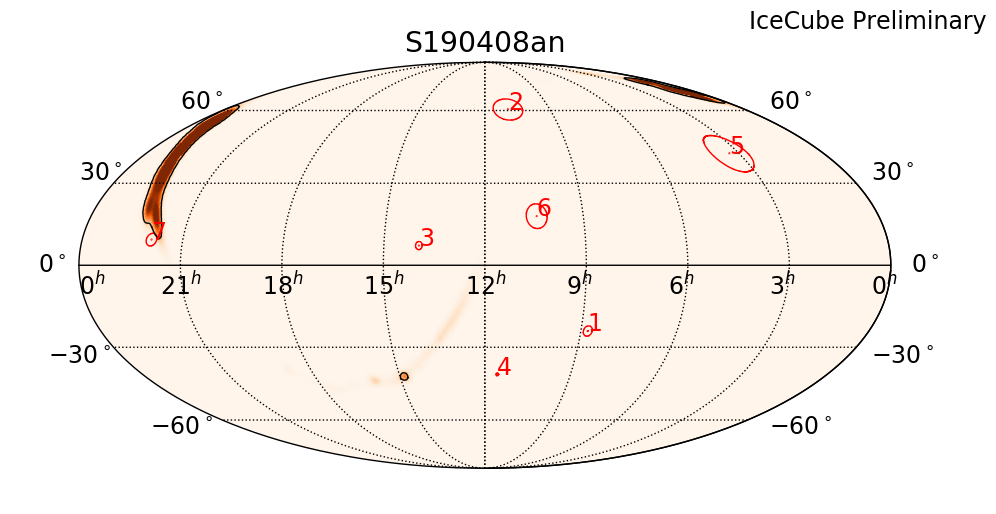} &
    \includegraphics[width=0.47\textwidth, trim=0cm 0.5cm 0cm 0cm]{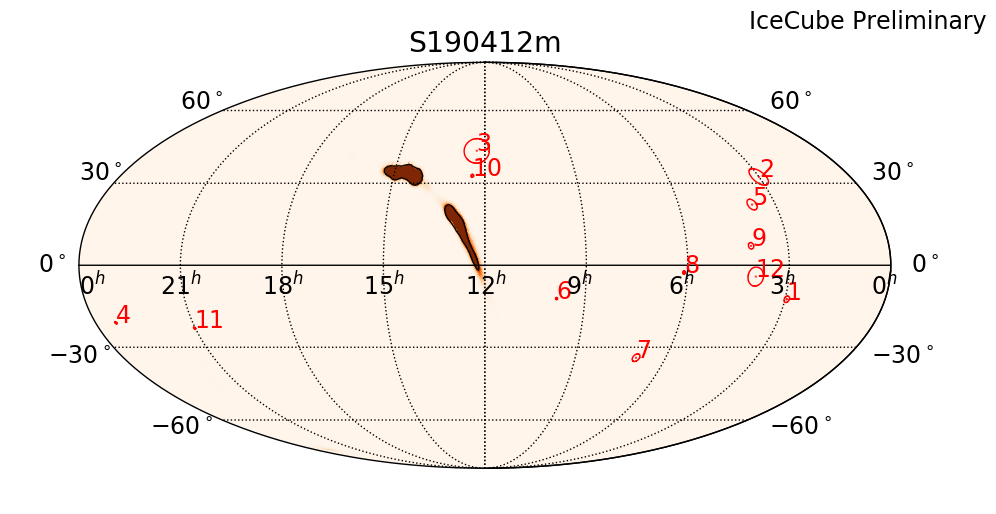} \\
    \includegraphics[width=0.47\textwidth, trim=0cm 0cm 0cm 0.5cm]{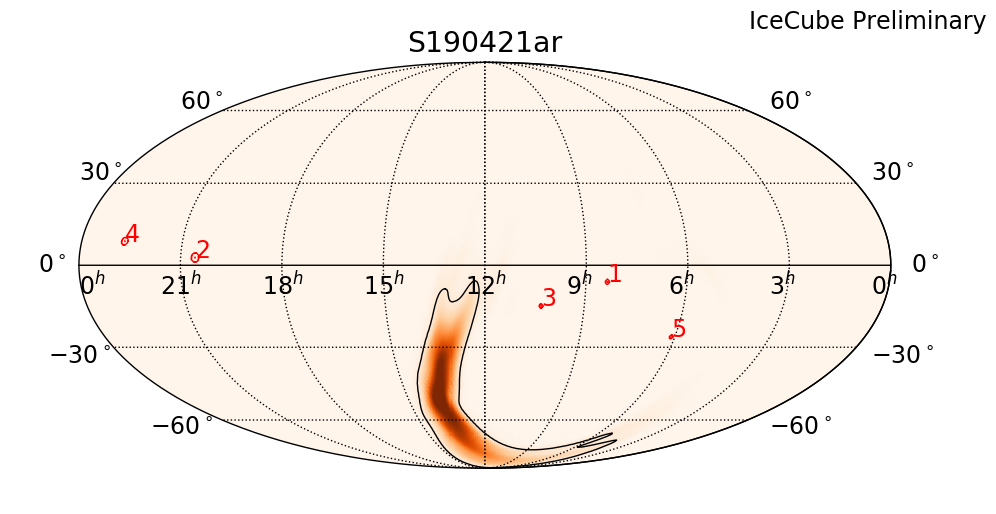} &
    \includegraphics[width=0.47\textwidth, trim=0cm 0cm 0cm 0.5cm]{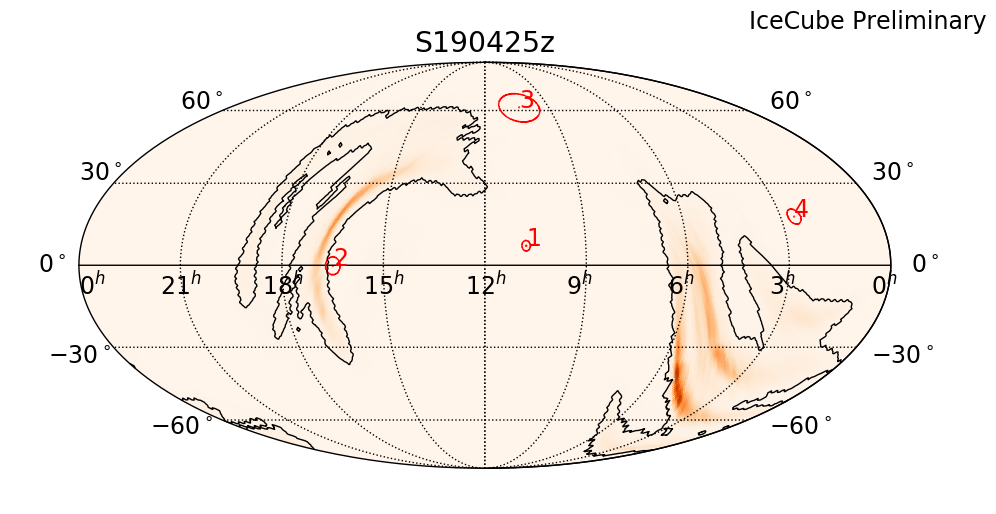} \\
    \includegraphics[width=0.47\textwidth, trim=0cm 0cm 0cm 0.5cm]{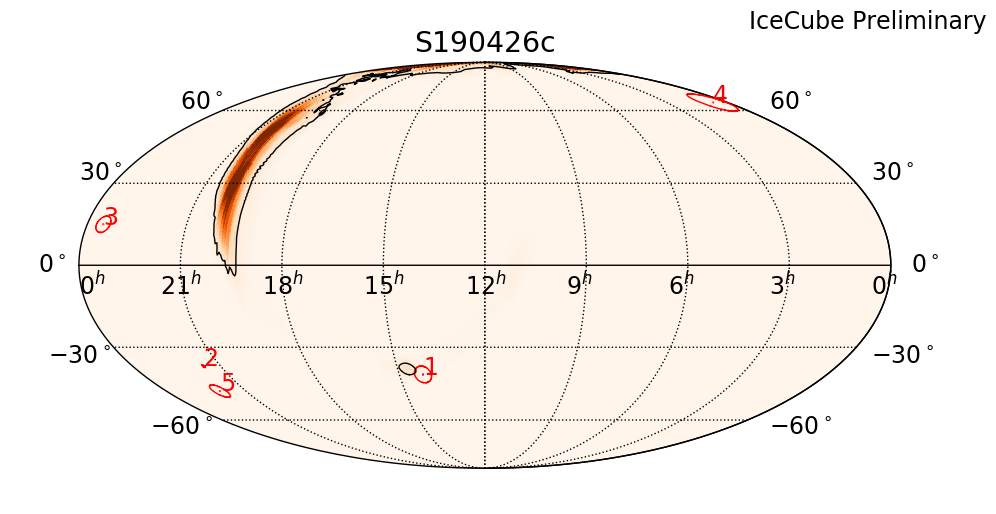} &
    \includegraphics[width=0.47\textwidth, trim=0cm 0cm 0cm 0.5cm]{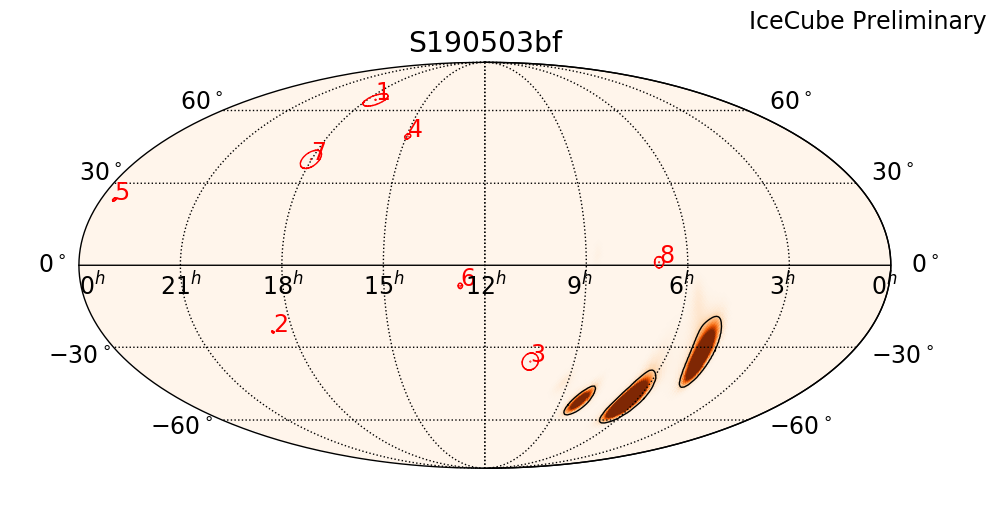} \\
    \includegraphics[width=0.47\textwidth, trim=0cm 0cm 0cm 0.5cm]{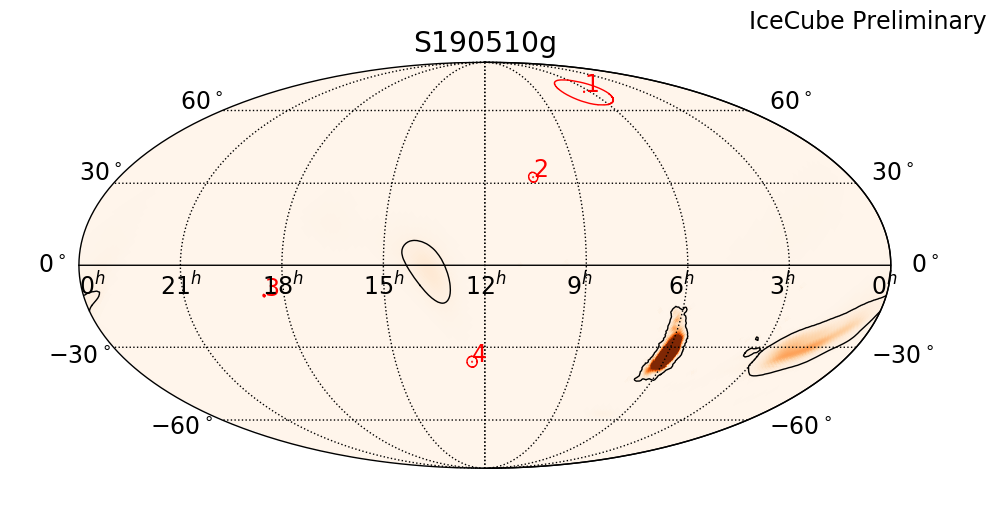} &
    \includegraphics[width=0.47\textwidth, trim=0cm 0cm 0cm 0.5cm]{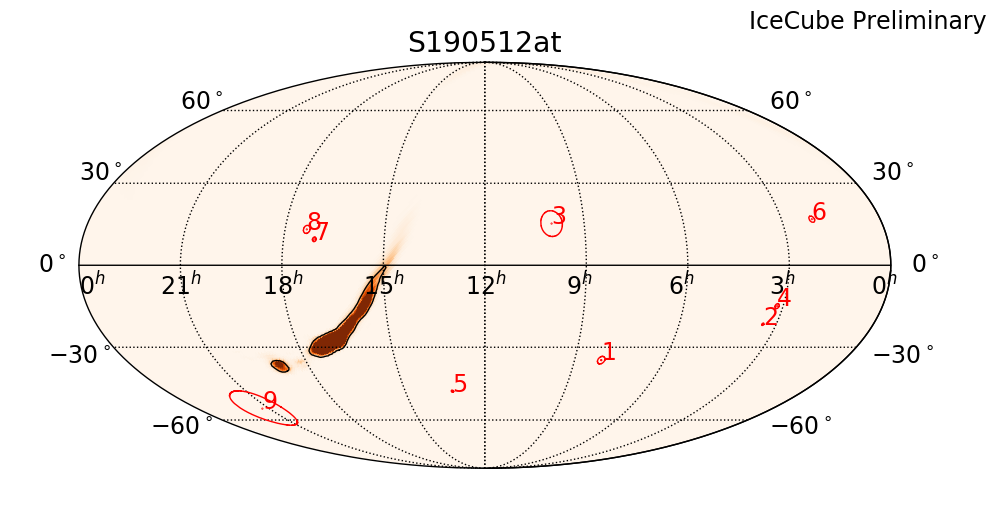} \\
    \includegraphics[width=0.47\linewidth, trim=0cm 0.5cm 0cm 0cm]{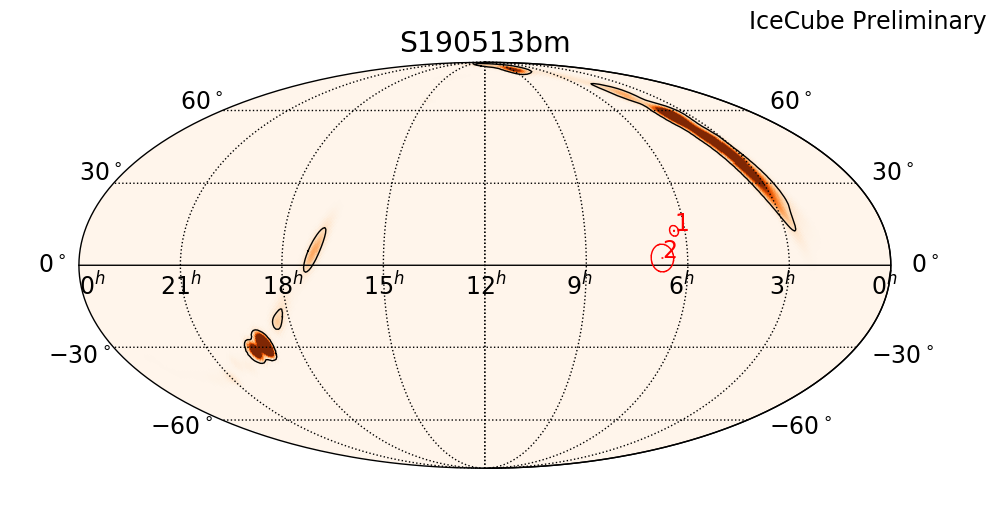} &
    \includegraphics[width=0.47\linewidth, trim=0cm 0.5cm 0cm 0cm]{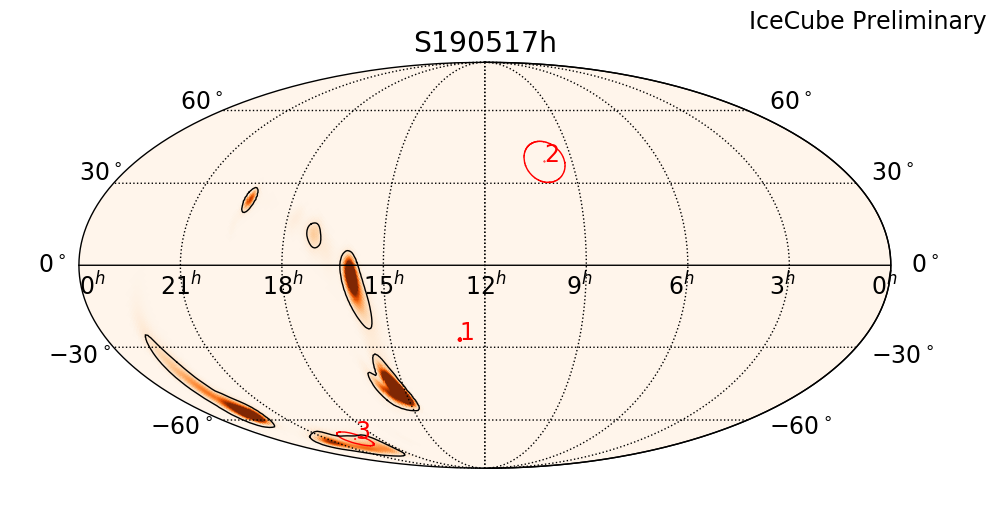}\\
\end{tabular}
\end{center}

\section{Search Method}\label{sec:method}
We use a Bayesian approach for the joint GW+HEN search~\cite{gwhenmethod}.
Our framework incorporates astrophysical priors of both types of messengers and their corresponding source as well as detector characteristics.
We define and compare multiple hypotheses for the joint GW+HEN event. 
Our signal hypothesis (H$_{\rm s}$) is that both GW event and at least one of the correlated HENs originated from the same astrophysical source. 
Our null hypothesis (H$_{\rm 0}$) is that triggers in both messengers arise from the background. 
Our chance coincidence hypothesis (H$_{\rm c}$) is that one type of messengers has an astrophysical trigger, but the other is a background event.
For the case of GW public alerts, H$_{\rm 0}$ is zero, because the GWs are considered to be real astrophysical events.

For GWs we use the following observational information for the search: (i) detection time, (ii) skymap which is reconstructed sky location probability density, (iii) signal-to-noise ratio (SNR), which is a measure of the event's consistency with background expectations, and (iv) reconstructed distance distribution. 
We define a vector containing the measured properties of a GW trigger as ${\bf x}_{\rm gw}$.
For HENs, we use: (i) detection time, (ii) reconstructed sky location probability density, and (iii) reconstructed energy proxy: the observed energy of the muon produced in the neutrino interaction in ice.
We assume that the reconstructed sky location can be described as a Gaussian distribution centered at reconstructed neutrino direction, with its reconstructed angular uncertainty. We define a matrix containing the measured properties of all neutrino triggers as ${\bf x}_{\rm \nu}$, with each row describing one neutrino.
We also define a vector containing our model parameters for the signal hypothesis as
$\theta=\{t_{\rm s}, r, \Omega, E_{\rm gw}, E_{\rm \nu}\},$
\noindent where $t_{\rm s}$ is the reference time, $r$ is the luminosity distance, $\Omega$ is the sky location, and $E_{\rm gw}$ and $E_{\nu}$ are the isotropic-equivalent total energies of GW and HEN, respectively.
Given the observational data, we compute a Bayes' factor for our signal hypothesis as 
\begin{equation}
\label{oddsratio}
O_{\rm gw+\nu} = \frac{P({\rm H}_{\rm s}|{\bf x}_{\rm gw}, {\bf x}_{\rm \nu})}{P({\rm H}_{\rm o}|{\bf x}_{\rm gw}, {\bf x}_{\rm \nu}) + P({\rm H}_{\rm c}|{\bf x}_{\rm gw}, {\bf x}_{\rm \nu})}.
\end{equation}
The detailed derivation of the Bayes' factor is explained in \cite{gwhenmethod}. 
Here we only explain a few important points about the priors that are used in calculating each term in Eq.~\ref{oddsratio}. 

To calculate the prior probability distribution of the source parameters, $P(\theta|H)$, we make a few assumptions: 
(i) time: a signal is equally likely to happen at any time during an observational period. Observation time is also independent of other parameters. Its prior probability distribution is therefore 
inverse of the livetime duration of the joint observation.
Although a time window of 1000~s is used in this search, any other period can be used for the search. 
When putting all terms together the final result of Bayes' factor is independent of livetime and it does not affect the significance of an astrophysical event. 
(ii) Distance: sources are uniformly distributed in the volume. 
(iii) Sky position: a uniform prior distribution in the sky is considered. 
(iv) Energy: independent log-uniform distributions are used for the energy ranges of GWs and HENs. 
To calculate the signal probability given the source parameters, we also consider the expected number of multi-messenger detections by estimating the expected number of detected HENs for given emission energy, sky location, and distance and using Poisson probability (equations 17-20 in \cite{gwhenmethod}).

Using Eq.~\ref{oddsratio}, we test our signal hypothesis. 
In order to convert this number to a frequentist significance, we use background data and simulations to empirically characterize the required threshold values. 
We first pick a representative of astrophysical source population from a set of injected parameters: mass, spin, sky position, inclination, distance, and arrival time.
We calculate SNR for each set of injection, considering a network of at least two GW detectors (the most sensitive case is with both of LIGO detectors and the Virgo detector). 
This way, we generate an ensemble of GW skymaps. 
Using an ensemble of scrambled GFU HENs, we calculate the Bayes' factor and the test statistic (TS) value for all coincident HENs with any GW event.
This provides a TS distribution that is used for characterizing the required threshold values. 
For any new GW event, TS can be calculated for HENs within $\pm 500$~s of GW detected time.
Comparing the derived TS value with the background TS distribution, a p-value is calculated for the combined HENs, which is the probability of a random GW+HEN background event having a TS greater than any given GW+HEN event. 

\section{Joint Alerts}

The LLAMA analysis ran in realtime during LIGO-Virgo's second observing run (O2; see e.g.~\cite{2016PhRvD..93l2010A}). 
The updated analysis discussed here has been running on GW realtime public alerts\footnote{https://emfollow.docs.ligo.org/userguide/} since the beginning of O3. 
At the time of writing this article, there have been 14 GW events; for all of which, the GW+HEN search ran through LLAMA pipeline. 
The results of these searches have been summarized in Table~\ref{tab:gwheno3} as well as Figure~\ref{fig:skymap}.
The p-values are reported for all HENs combined in each search. 
No event with a p-value~<~1\% have been found to date. 
The p-values for BNS candidates are not reported here due to currently revisiting the BNS distributions for this analysis. 

\begin{table}
\begin{center}
\begin{tabularx}{\linewidth}{|p{0.4cm}|p{1.63cm}|L|p{3.68cm}|} 
\hline
No. & GW event & Possible Source (probability) & p-value (binary merger) [preliminary] \\
\hline
1 & S190408an & BBH (>99\%) & 0.15 \\
\hline
2 & S190412m & BBH (>99\%) & 0.83 \\
\hline
3 & S190421ar & BBH (97\%), Trs (3\%) & 0.62 \\
\hline
4 & S190425z & BNS (>99\%) & --\\
\hline
5 & S190426c & BNS (49\%), NSBH (13\%), Trs (14\%), MG (24\%) & -- \\
\hline
6 & S190503bf & BBH (96\%), MG (3\%) & 0.29 \\
\hline
7 & S190510g & BNS (42\%), Trs (58\%) & -- \\
\hline
8 & S190512at & BBH (99\%), Trs (1\%) & 0.51 \\
\hline
9 & S190513bm & BBH (94\%), MG (5\%) & 0.74 \\
\hline
10 & S190517h & BBH (98\%), MG (2\%) & 0.12 \\
\hline
11 & S190519bj & BBH (96\%), Trs (4\%) & 0.16 \\
\hline
12 & S190521g & BBH (97\%), Trs (3\%) & 0.19 \\
\hline
13 & S190521r & BBH (>99\%) & 0.16 \\
\hline
14 & S190602aq & BBH (>99\%) & 0.13 \\
\hline
\end{tabularx}
\end{center}
\caption{GW events detected so far during O3, the possible source classification including their probabilities, and the p-values of the LLAMA GW+HEN search. 
The possible source classifications are Binary Black Hole (BBH), BNS, Terrestrial (Trs), or MassGap (MG; i.e. at least one compact object in a binary system has a mass 3-5 solar masses).
The p-values for BBH candidates are preliminary and are not reported for BNS candidates as the BNS distributions are currently being revisited for this analysis.}
\label{tab:gwheno3}
\end{table}

Two different GCN circular templates will be used in response to the GW alerts, depending on the p-values reported by the two GW+HEN searches. 
The first case is when neither of the searches report a p-value~$ <1\%$. 
The GCN circular in this case will report a non-detection as well as fluence upper limits/sensitivity ranges.
The second case is either or both analyses find a p-value~$ <1\%$. 
In this case, we will report directions (right ascension and declination), angular uncertainties of neutrino event (the radius of a circle representing 90\% confidence level containment by area), time offset of neutrino with respect to GW trigger, and p-values from both analyses for all significant neutrino events found in coincident with the GW candidate. 
The two p-values will be different due to testing different hypotheses:
The LLAMA search presented here uses a Bayesian approach to quantify the joint GW+HEN event significance, 
and accounting for known astrophysical priors in the significance estimate, such as GW source distance.
The other~\cite{Hussain:2019icrc_gw} is a maximum likelihood analysis which searches for a generic point-like source coincident with the given GW skymap. 

\section{Discovery Potential}
\begin{figure}
    \begin{center}
        \resizebox{0.7\textwidth}{!}{\includegraphics[trim=0cm 0.5cm 0cm 0.5cm]{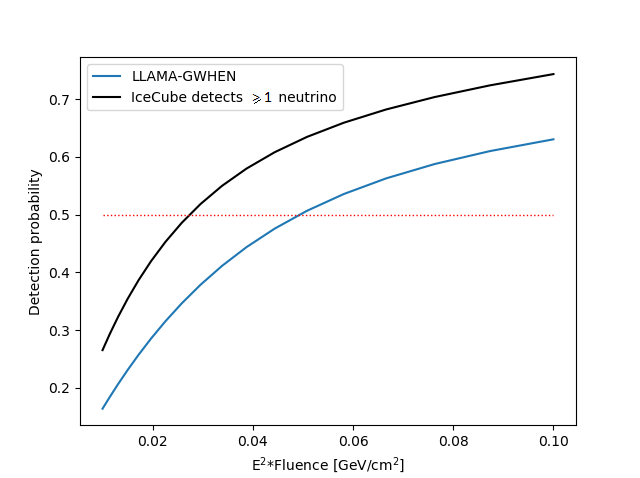}}
    \end{center}
    \caption{Discovery potential of the LLAMA search for GW+HEN events vs neutrino fluence (blue). A fluence of 0.049  GeV/cm$^2$ is detected with 50\%, averaged of the entire sky.  The fluence that results in at least one neutrino observation by IceCube is also plotted (black). The discovery potential is defined as the signal strength that leads to a 3$\sigma$ deviation from background in 50\% of all cases. The fluence is reported assuming an E$^{-2}$ energy spectrum.}
    \label{fig:disc}
\end{figure}

To test the performance of LLAMA in searching for GW+HEN joint events, we express the discovery potential in terms of fluence, defined for an E$^{-2}$ spectrum. 
The discovery potential is defined as the signal strength that leads to a 3$\sigma$ deviation from background in 50\% of all cases. 
Figure~\ref{fig:disc} shows the detection probability versus fluence. The discovery potential of the search is indicated by the point where the red dotted line intersects with the blue curve. 
For comparison, the probability that IceCube detects a single neutrino at each fluence is plotted.
Averaged on the entire sky, a fluence of 0.049  GeV/cm$^2$ is detected with 50\% probability.

\section{Future Prospects}
Joint searches of GW and HEN can help us better understand the interaction of relativistic and sub-relativistic outflows from BNS mergers. 
Even in cases that EM signals are not detectable, we can still probe the jet physics with detected HENs.
For example, in the case of GW170817, $\gamma$-rays were marginally detected with \textit{Fermi}-GBM satellite and if the source was more distant, $\gamma$-rays would have not been detected. 
In these situations, neutrinos from the BNS merger might still be detectable by IceCube~\cite{shigeo2018} and provide useful insights into studying jet physics. 

EM counterparts of GW170817 event confirmed that considerable amount of matter is expected to be ejected from BNS mergers (see~\cite{shigeo2018} and references therein).
This matter may either produce a successful jet or choke the relativistic jet. 
Ref.~\cite{shigeo2018} discusses in detail two models that can produce neutrinos from BNS mergers for the situation that the jet is choked inside the kilonova ejecta.
In this case, the EM emission might be completely absorbed by the ejecta and the neutrinos that are produced inside the ejecta are the only messengers that are able to escape the remnant and be detected on the ground which would enable us to probe the choked jet physics.  

The non-detection of neutrinos from BNS mergers can place limits on the choked jet parameters. 
In the case of GW170817, the event was followed by a faint sGRB, 
a kilonova, and a slowly brightening afterglow, which could indicate that the jet was viewed off-axis. 
Alternatively, a wide-angle cocoon could be inflated by the choked jet which could produce HENs. 
We should note that the effective area of IceCube is smaller in the southern sky where GW170817 appeared which could be a reason for non-detection of neutrinos from this event.
In case a BNS merger happens in the northern sky where IceCube has lower background events and higher effective area, it is possible to either detect neutrino events or place limits on the models such as choked jet parameters.

\section{Conclusion}\label{sec:conc}

We have developed a realtime pipeline called LLAMA to search for significant coincidences between GWs and HENs.  
We use data from IceCube's GFU stream and public alerts of LIGO/Virgo for this search.
A Bayesian analysis considering astrophysical priors and detector characteristics is developed and used as our TS to find a significance for each joint coincident GW+HEN event.
For any GW trigger, we report the results of LLAMA search along with the results of the other search~\cite{Hussain:2019icrc_gw} in a GCN circular.
We have reported the results of our searches since the start of O3 in this paper, with no GW+HEN event with p-value~<~1\% observed so far. 
Joint GW+HEN searches may help us better understand the physics of GW+HEN astrophysical sources.

\noindent{\bf Acknowledgements.} AK thanks Columbia University Frontiers of Science fellowship. The authors thank the University of Florida and Columbia University in the City of New York for their generous support.
The Columbia Experimental Gravity group is grateful for the generous support of the National Science Foundation under grant PHY-1708028.


\bibliographystyle{ICRC}
\bibliography{keivani-icrc19}

%

\end{document}